\documentclass[iop]{emulateapj}
\usepackage{graphics}
 \usepackage{epsfig}  
\usepackage{psfig}

\def\gtrsim{\mathrel{\hbox{\rlap{\hbox{\lower4pt\hbox{$\sim$}}}\hbox{$>$}}}}
\def\ltsim{\mathrel{\hbox{\rlap{\hbox{\lower4pt\hbox{$\sim$}}}\hbox{$<$}}}}

\begin{document} 

\newcommand{\fuse}{{\it FUSE}} 

\newcommand{\an}{\,\AA}

\title{A Detailed Far-Ultraviolet Spectral Atlas of O-Type Stars }

\author{Myron A. Smith} 
\affil{Catholic University of America,\\
        3700 San Martin Dr.,
        Baltimore, MD 21218;           \\
        msmith@stsci.edu}

\begin{abstract}
  In this paper we present a spectral atlas covering the wavelength interval 
930--1188\,\an\ for O2--O9.5 stars using {\it Far Ultraviolet Spectroscopic 
Explorer} archival data.  The stars selected for the atlas were drawn from 
three populations: Galactic main sequence (class III-V) stars, supergiants, and
main sequence stars in the Magellanic Clouds, which have low metallicities.
For each of these stars we have prepared FITS files comprised of pairs of 
merged spectra for user access via the {\it Multi-Mission Archives 
at Space Telescope.}
We chose spectra from the first population with spectral types O4, O5, O6, O7, 
O8, and O9.5 and used them to compile tables and figures with identifications 
of all possible atmospheric and ISM lines in the region 949-1188\,\an.~  
Our identified line totals for these six representative spectra are
821 (500), 992 (663), 1077 (749), 1178 (847), 1359 (1001), and 1798 (1392)
lines, respectively, where the numbers in parentheses are the totals of 
lines formed in the atmospheres, according to spectral synthesis models. 
The total number of unique atmospheric identifications for the 
six main sequence O star template spectra is 1792, whereas the 
number of atmospheric lines in common to these spectra is 300.
The number of identified lines decreases toward earlier types 
(increasing effective temperature), while the
percentages of ``missed" features (lines not predicted from our
spectral syntheses) drops from a high of 8\% at type B0.2, from our
recently published B star far-UV atlas, to 1--3\% for type O spectra.
The percentages of overpredicted lines are similar, despite their being
much higher for B star spectra. We discuss the statistics of line
populations among the various elemental ionization states. Also, as an
aid to users we list those isolated lines that can be used to determine 
stellar temperatures and the presence of possible chemical anomalies. 
Finally, we have prepared FITS files giving pairs of merged spectra for stars 
in our population sequences for access via the {\it Multi-Mission 
Archives at Space Telescope.} 
\end{abstract}

\clearpage

\section{Introduction}
\label{int}

High dispersion spectroscopy provides windows into the past and current
physical processes in massive O stars as clear as for stars anywhere on the 
H-R Diagram.  Individual O stars can be found in different stages of evolution
because of their short lifetimes and unique spectral signatures that advertise
these stages. 
Indeed, many of them evolve to SN\,Ib and SN\,Ic supernovae by first becoming 
Luminous Blue Variables and/or Wolf-Rayet stars, which are easily 
identifiable by their light curves or broad spectral emission features. 
Likewise, O stars in close binaries are readily discovered from emission
in optical and UV lines excited by interacting winds. 
Because the incidence of lines in O star spectra and continuum flux 
both peak in the far-ultraviolet wavelength region, the use of space-borne
instrumentation is critical to fully understanding their atmospheres and 
radiative processes.
Although observations with the {\it Far Ultraviolet Spectroscopic Explorer} 
satellite (\fuse) were terminated in 2007, the \fuse\ archive now offers 
a large sample of stellar spectra having identical wavelength coverages 
provided by homogeneous data processing.
In particular, \fuse-accessible O stars reside not only within the solar 
neighborhood in the Galaxy but also in the more distant Magellanic Clouds. 
Because their lifetimes are brief, O stars provide an 
invaluable record of the recent chemical composition and angular
momentum histories in the solar neighborhood and the return of much of 
their nuclear-processed matter to the Interstellar Medium (ISM).

  Astronomers have historically been quick to capitalize on uniformly 
reprocessed data held in archives following the close of UV spectroscopic 
mission.  The availability of these archives has allowed the construction
of a number of fine spectral atlases recorded by the {\it Copernicus} and 
the {\it International Ultraviolet Explorer ( IUE}) satellites.
Important examples of these compilations for the middle-UV spectral range are 
the {\it Copernicus} atlas of $\tau$\,Sco (Rogerson et al. 1978) and {\it IUE} 
pictorial atlases of O and B stars (Walborn, Nichols-Bohlin, \& Panek 1985, 
Rountree \& Sonneborn 1993, Walborn, Parker, \& Nichols 1995).
Pellerin et al. (2002) inaugurated a second group of digitized far-UV (\fuse) 
atlases consisting of representative spectra of O and early-B type, 
luminosity class I-V stars in the Galaxy and Magellanic Clouds.
These atlases have been supplemented by figures
constructed by N. Walborn for the Gray \& Corbally (2009) monograph on stellar
spectral classification that show the changes of primary spectral features 
in the far-/middle-UV with luminosity class in main sequence O stars. 

  Such atlases have been effective in showcasing the general trends of the
strong photospheric lines and the so-called ``UV resonance wind" lines 
with effective temperature T$_{\rm eff}$ and luminosity class. For example,
a reconnaissance of the O-star (Pellerin et. al) atlas shows how far-UV 
spectra of most O stars are strongly mutilated by interstellar lines,
particularly below 1100\,\an. The photospheric component of these 
spectra is dominated by the confluence of high series Lyman lines, 
Ly$\beta$--Ly$\theta$ down to the blue limit of the \fuse\ instrument 
at 920\,\an.~ The strongest metallic feature formed substantially in the 
photosphere is generally the C\,III line complex at 1174--1176\,\an. 
This multiplet arises from an excitation of the lower atomic levels at an
excitation  $\chi$$_{exc}$ = 9\,eV.~ In supergiant O stars radiative winds 
can cause this complex to develop into a broad P\,Cygni profile.
Other strong features are resonance lines of 
C\,III (977\,\an), N\,III (991\,\an), 
O\,VI (1031\,\an, ~1037\,\an), Si\,IV (1128\,\an), S\,IV (1062\,\an,~ 
1072\,\an,~ 1073\,\an,~ 1098--1100\,\an), P\,IV (950\,\an), and 
P\,V (1117\,\an,~ 1128\,\an). Spectral lines of O stars suffer doppler
broadening, and this serves to wash out isolated weak metallic lines
that would otherwise be resolved in high dispersion spectra.
However, as detailed below, the far-UV spectra have the saving grace that, 
except for the liberally distributed ISM lines, the photospheric lines tend
to be more widely spaced and to blend less with one another than in the
middle-UV wavelengths.

The Pellerin et al. atlas and Barnstedt et al. (2000) 
have displayed valuable detailed information about
the presence of molecular H$_2$ features formed in the ISM. From the point
of view of stellar atmosphere investigators, these features contaminate 
many of the far-UV lines formed in the atmospheres of most Galactic stars. 
The Pellerin et al. atlas was closely followed by a second spectral atlas of 
OB stars in the two Magellanic Clouds (Walborn et al. 2002).  This work 
concentrated on the behavior of wind lines with respect to metallicity as 
well as effective temperature and luminosity.  In addition, Blair et al.
(2009) published a compendium of {\it FUSE} spectra of hot stars in the
Magellanic Clouds. This atlas focused on the identification of far-UV 
resonance lines formed in the ISM, including those found at 
multiple velocities. 

  The first far-UV high-dispersion spectral coverage published for a
B star was the {\it Copernicus} atlas of $\tau$\,Scorpii (B0.2\,V;
Walborn 1971) by Rogerson \& Upson (1978).
Rogerson \& Ewell (1985; ``RE") published a detailed tabulation
of atmospheric and ISM lines identified in this atlas.
The RE work was undertaken at
a time when spectral synthesis tools were not commonly available,
and when only a relative handful of experienced spectroscopists who were
also specialists in atomic physics could make reliable line identifications.
Even so, in the absence of commonly available 
synthesis programs at that time, it was difficult to make wholesale
line identifications without some errors. This
situation has changed dramatically in the intervening years with
the development of spectral line synthesis tools that make use
of extensive atomic line libraries. 

 Inspired by the {\it Copernicus} atlas, Smith (2010; hereafter ``Paper\,1") 
constructed a far-UV spectral atlas for B stars using spectra from \fuse\, 
{\it IUE,} and the {\it Space Telescope Imaging Spectrograph} ({\it STIS}) 
over the (vacuum) wavelength range 930--1225\,\AA.~ 
In making this atlas we set an arbitrary short wavelength limit of 930\,\an\ 
because, other than for a few blended Lyman lines, no useful information
about the photospheric spectrum could be obtained below this wavelength. 
We chose the red limit by including spectra from the {\it HST/STIS} 
and/or {\it IUE} cameras in order to cover the Lyman\,$\alpha$ 
feature and other important lines in its vicinity.  The B star atlas 
addressed its first goal of fulfilling the need for the identification of all 
possible visible lines in the ``template" spectra of three representative 
main sequence B0.2\,V, B2\,V, and B8\,V stars. These identifications were
made by using published atomic line libraries that predict occurrences 
of lines from spectral synthesis models. This atlas also recorded ``misses," 
i.e., the wavelengths of observed lines that could not be predicted from our 
line syntheses.  The rotational velocities of stars selected for the atlas 
are relatively low in order to resolve as many neighboring photospheric lines
as possible. Therefore, we referred to this work as a ``detailed spectral
atlas." The second goal of this atlas was to provide uniform spectral data 
products to the astronomical community through the wavelength range just
described.

The numbers of lines identified for the three template spectra in Paper\,1 
were 2288 (2004), 1612 (1465), and 2469 (2260) lines, respectively. 
Here the values given in parentheses are the number of lines found within 
almost the full wavelength range covered by the \fuse\, 949--1188\,\an.~
Of these totals only small percentages (8\%, 2\%, and 2\%, respectively) 
of photospheric lines could not be identified, which is to
say that the oscillator strengths (log\,$gf$'s) for these lines are
at best poorly determined and are thus not included in our line library. 
The B star atlas results were published more fully in the electronic edition, 
The spectral data files and all other products were made available for
public download as one of MAST's\footnote{The Multimission Archives for Space 
Telescopes is located at the Space Telescope Science Institute (STScI).
STScI is operated by the Association of Universities for Research in
Astronomy, Inc., under NASA contract.  Support for 
archiving MAST data is provided by NASA Office of Space Science under grant 
NAS5-7584.} ``High Level Science Product" (HLSP) area
(http://archive.stsci.edu/prepds/fuvbstars/). 

   This paper presents a similar spectral atlas for O stars. The atlas 
is organized in much the same way as the B star atlas. However, one 
important difference is that whereas the blue wavelength limit we chose, 
930\an, is the same as for the B star atlas, the red one is set by
the \fuse\ spectral coverage, again, $\approx$1188\an. In particular, 
we found for the present atlas that it was not easy to again include the 
1188--1225\an\ region that was previously surveyed in the B star atlas because 
of the paucity of O stars observed systematically in this spectral region.

Our presentation is organized as follows. The selection of spectra and
the methodology for data handling, including the creation of FITS spectra
for all stars in our sample, as well as for the line identification in six
exemplars, are described in $\S$2. In $\S$3 we display portions of the
atlas and give a detailed list of several thousand 
identifications as well as a list of ``clean" lines across
much of the O-star domain.
In $\S$4 we give relevant statistics from our identifications for 
the six O4\,V, O5\,V, O6.5\,V, O7\,III, O8\,V, and O9.5\,III
exemplars we call our O star spectral templates.
We also comment in detail on a number of possible spectral
markers for physical conditions (mainly effective temperatures)
in these stars' atmospheres, all of which are assumed to be in
hydrostatic equilibrium.

\section{Preparation of the Atlas}
\label{obsn}

\subsection{Atlas star selection} 
\label{opt}

  The selection of stars for our atlas was based on a number of criteria.
These included high signal-to-noise ratios (SNR), absence (so far 
as is known) of double-lined binary signatures, low to
moderate rotational broadening ($v\,sin\,i$ $\ltsim$ 100\,km\,s$^{-1}$), 
and far-UV extinction ($E(B-V) <$ 0.4) from the ISM. 
An additional criterion was that their surface iron group metallicities should 
be close to those of most other Galactic young massive stars in the solar 
neighborhood. We also wanted to augment the scope of atlas by including 
supergiants, even though their lines are usually affected by rotational and
atmospheric doppler broadening mechanisms.
While all these criteria are desirable for a spectral atlas of 
massive hot stars in the local region of the Galaxy, they can conflict
with one another at times. A second issue affecting our our selections 
is that those stars with solar-like compositions are strongly confined 
to the Galactic plane and suffer extensive contamination by molecular H$_2$ 
and atomic ISM features as well as attenuation from ISM dust. This
absorption peaks in the far-UV region. 
To include representative O stars beyond the plane that exhibit smaller
amounts of far-UV absorption, one is generally obliged to 
find them in one of the Magellanic Clouds, 
and these stars have low metallicities.
Altogether, we included parallel 
sequences of main sequence (luminosity class III-V) stars observed at 
high galactic latitudes (members of the Magellanic Clouds)
as well as a third sequence of supergiants (class I--II). 

  A survey of the O star data in the entire MAST/\fuse\ archive demonstrated 
that we could not fill a table of representative stars across the O domain 
using all these sometimes conflicting criteria. In particular, we could 
not find spectra in the archive of stars having spectral types of O2 and O3 
and supergiant luminosity classes.  Otherwise, we were able to find three 
stars per spectral type over the range O4--O9.5, giving us a total of 25 stars. 
These stars, their spectral types, and $v\,sin\,i$ values, and 
references for these values are listed in Table\,1. 
We took most of these spectral types from the compilations 
of Howarth et al. (1997; ``H97") and Penny \& Gies (2009; ``PG09").
We have grouped names of three stars for spectral type (excepting O2 and O3) 
in order of the three stellar populations listed just above.

Stars marked by asterisks in the table represent the six main sequence
template stars referred to above.
For our line identification tables for six main sequence stars we assign 
``star code" values of 1--6 to those stars in our line identification tables. 
In our system the value
1 corresponds to subtype O9\,5, 2 to O8 (omitting O9), and so on, until the 
value 6 is applied to O4.  Note in addition that we were not able to find 
high quality spectra of examples of O2 and O3 supergiants.

\subsection{Data properties and handling} 

\subsubsection{Conditioning of {\it FUSE} spectra}

  During its eight year lifetime the {\it FUSE} spacecraft was operated with 
four independent spectrographs/cameras.  As described in the {\it FUSE Archival 
Instrument Handbook} (Kaiser \& Kruk 2009), light collected from each of the 
telescopes illuminated a holographic diffraction grating/camera mounted on 
a Rowland circle spectrograph. LiF and SiC coated camera mirrors focused 
the dispersed light on two far-UV microchannel plate detectors.
Table\,2 lists in italics the (vacuum) wavelength coverage of each of 
the {\it FUSE} detector segments.
As noted above, nearly complete far-UV coverage of the spectrum was 
provided by two nearly identical ``Sides" 1 and 2 of the instrument. 
Each of these Sides included two pairs of LiF and SiC detectors.  
The parenthesized wavelengths intervals in the table give the regions for 
each Side/detector combination. Coverage of each wavelength was provided
by each of the two Sides, with two exceptions:
(1) data for the wavelength interval 
1090--1094.5\,\an\ were recorded only from Side\,2 detectors, 
and (2) data for 1015--1075\,\an\ were recorded by all four detectors.
The spectral resolution (full width half maxima)
of \fuse\ spectra is 15-20 km\,s$^{-1}$, i.e., 0.05--0.07\,\an, 
though this varies slightly from one detector segment to another. 
Spectral fluxes were sampled by the pipeline processing system at uniform
intervals of 0.013\,\an~, and we maintained this spacing in our products. 

The conditioning steps needed to produce continuous spectra from \fuse\ 
archival data are somewhat complex. First, the wavelength calibrations 
for \fuse\ spectra are known to undergo excursions from a linear dispersion
of up to $\pm{0.03}$\,\an\ over intervals as short as 5 \an ngstroms.
These excursions are largely due to electron repulsion in the detector 
that distort the positions of recorded photoelectrons in the 
detector $x$ and $y$ axes. The effects are typically not robust with time 
and therefore cannot be modeled or compensated for reliably.
In addition, small wavelength shifts due to positioning and
wandering of the star image in the science apertures are the norm. 
Since most of our spectra were obtained with multiple exposures, the
first step in correcting them was to cross-correlate them {\it inter alia}
to introduce subpixel shifts to the system of the first observations. 
The shifted spectra were manually inspected to insure that this step was 
performed accurately. We then coadded the spectra according to the 
pipeline-generated errors for the individual exposures. 
To remove any residual offset shifts and local departures from 
the mean dispersion, we wrote a program in the {\it Interactive Data
Language}  to view and match the observed features of the Side 1 and Side 2
spectra with a comb spectrum  of H$_2$ features provided by the online 
``H2ools" tool (McCandliss 2003). 
The features in H2ools matched the computed Lyman and 
Werner transitions from a single rotational state of the zeroth or 
occasionally first vibrational level to the ground state of molecular H$_2$ 
lines.  It is important to point out one key difference between ISM and 
photospheric line identifications:  whereas the photospheric lines are 
based on {\it predictions} from the line synthesis spectra for an assumed
T$_{\rm eff}$ of a spectral type, all ISM identifications are based on their
{\it visibility} in the template spectra. This could mean, for example, 
that a given H$_2$ feature may not be found in a particular template
spectrum if the column density toward the star referred to is smaller
than toward other stars.

To reference the absolute wavelength system from the H$_2$ features we 
had to rely on estimates of the locations of strong unblended photospheric
lines relative to nearby H$_2$ features. 
The doppler shifts of our Galactic stars are typically within $\pm{0.05}$\an. 
Exceptions to this statement are for our O8\,V line identification star 
HD\,66788 and the O9\,Ib-II star HD\,207198. The HD\,66788 spectrum is an 
unusual case because it relies on a single \fuse\ exposure. The spectrum 
showed a wavelength shift of +0.6\,\an\ ($\approx$85 km\,s$^{-1}$) from
the Galactic system set by the H$_2$ lines.
The dataset for HD\,207198, which is known to be in a spectroscopic binary, 
is comprised of five exposures over two epochs. Their shifts are
the same for each of two common epochs, but the archived spectra in one group
are shifted 4 pixels from the spectra of the other group. When referenced
to the first epoch, the averaged spectrum exhibits an additional shift of
+0.2\,\an\ with respect to the H$_2$ scale. Thus, both stars appear to be 
primaries of binary systems. However, we could not discern line 
components of a secondary star in the spectrum. The pipeline spectra of
all the Magellanic (AZ and Sk) objects are shifted
from the Galactic H$_2$ wavelength system by +0.6\,\an, except for the
O8\,V star D301-NW8, which is shifted by +1.2\,\an.
The spectral shifts from the nine observations suggest that this object is
the primary of a second undiscovered binary system in our star list. 
Such large wavelength shifts are handy to have in our sample because they
help to separate contributions of overlapping features 
formed in the photosphere and ISM. We have also made judicious use of this 
advantage by using the O8 template to arbitrate between identifications in 
other spectra that are unshifted from the rest (ISM H$_2$) frame.

  A final complication in {\it FUSE} spectra was the occurrence of optical
vignetting across the detector fields called ``worms." 
These cause recognizable flux depressions across
certain detector segments, especially for the LiF 1B segment in
the region 1130\,\an\ to 1160\,\an\ (see Chapter~4 of Kaiser \& Kruk 2009). 
We addressed this feature by passing a high order filter having the
same degree over the 1B and 2A segment fluxes and then interpolating 
to the original pixel scale. The depression in the Side 1B spectrum 
was mitigated by dividing the 
smoothed polynomial fit of the 2A spectrum by the similarly smooth fit of 
the 1B one in the wavelength region affected by the worm. The quotient 
spectrum was then applied to the original 1B spectrum.  A second worm in 
the LiF Side\,1 detector spectrum occurs in the region 997--998\,\an. However,
this worm was left uncorrected in the plots and spectral data discussed below. 
In addition to worms, we noticed that occasional drop offs of flux at the 
ends of segments, e.g., 1087--1088\,\an\ for SiC-B Side\,2. 
These occurrences appeared over too short a wavelength span to correct.
Otherwise, because several instrumental idiosyncracies appear in
{\it FUSE} detector Sides 1 and 2 spectra, most researchers have learned
to work with the segments of these two sides separately and to compare them
in the end as if they were independent observations. We have done the same
in preparing our atlas by merging the four segments of the two sides to
form two long spectra, with break points given in Table\,2.
Thus, the last step in establishing the wavelength calibration was to 
correct the spectra for the radial velocity shifts discussed above and 
to interpolate them to the initial grid from the pipeline system, that is, 
starting at 920.0\,\an\ and incrementing by 0.013\,\an\ to about 1188.4\,\an.
The spectra for the 25 stars in this atlas are given for the two independent 
detector sides as separate extensions in our products, which are Flexible 
Image Transport System (FITS) files, formatted as binary numbers.

\subsection{Tools for line identifications}

\subsubsection{Six template spectra}

  We expect the primary use of our atlas will come from the consultation of 
line identification tables and annotated plots discussed below. It is 
unnecessary and time-consuming to attempt to identify lines according for 
each of the 25 stars given in FITS format. Our philosophy is to identify 
all lines that can be predicted in the far-UV spectrum in increments of 
effective temperature of $\approx$2\,kK up to T$_{\rm eff}$ = 42\,kK. 
(Our model atmospheres line synthesis program discussed below operates 
on models computed with T$_{\rm eff}$ values in integral kilokelvins.) 
We chose six template representative spectra of metal-normal O dwarf 
stars for our line identifications. 
Our T$_{\rm eff}$ values in Table\,1 are for the most part those that follow
the spectral type calibration of Martins, Schaerer, \& Hillier (2005, their 
Table\,4). The exception to this practice is that we used T$_{\rm eff}$ = 
34\,kK (rather than 35\,kK) to the value that Marcolo et al. (2009) determined 
for our O8\,V star, HD\,66788. We note also that the value T$_{\rm eff}$ = 
32\,kK used for the O9.5\,V template spectrum is similarly spaced from the 
highest value, 30\,kK, that we used in Paper\,1 for identifications in the 
``anchor" ({\it Copernicus}) spectrum of $\tau$\,Sco.

\subsubsection{Construction of the line library}

To prepare for the identification of lines in our spectral
templates using spectral synthesis models, we compiled a line library
from three sources: the Kurucz (1993) line library, the Vienna
Atomic Line Database (``VALD"; Piskunov et al. 1995, Kupka et al.
1999), and the on-line atomic line database of van Hoof (2006).
The Kurucz line library is comprehensive and determined from atomic
theory computations. In the far-UV this provides the advantage 
that the library coverage is not compromised as it otherwise could be by 
absorptive optical coatings used in laboratory spectrographs. 
However, even though it continues to provide
most of the identifications we used, many of the oscillator strengths 
(log\,$gf$s) in this list have become dated. 
By contrast, the VALD and van Hoof databases are periodically updated. 
Both also give a recommended log\,$gf$ value for a line if more 
than one have been published. The VALD library is supported by an 
interactive web interface. This access tool allows a user to input on an 
web form a rough photospheric effective temperature and as well
as line depth threshold criterion, in our case 1\% line depth. All lines
computed with depths larger than the threshold were included in 
a returned list.
We exercised this option and chose parameters T$_{\rm eff}$ = 35\,kK and 
solar abundances in our requests to VALD of expected detectable lines through
our line identification -UV region, 949\,\an\ to 1188\,\an. The short limit 
was set as the starting point of the {\it Copernicus} atlas for $\tau$\,Sco.
We repeated the procedure with T$_{\rm eff}$ = 45\,kK.
Our requests to VALD using these two effective temperature resulted 
in the addition of new lines to our original B star line library. 
The van Hoof tool is convenient to use both in augmenting the atomic line list 
and for the spot checking we occasionally had to do to confirm our automated 
line synthesis results.

The final stage of the line selection was to screen out line duplications
that sometimes occur in combining lines from different library sources. 
We accomplished this by writing a program to run through our list and identify 
line entries within ${\pm 2}$ m\an\, arising 
from lower excitation levels within ${\pm 0.03}$\,eV of the same ion. 
In these cases lines with the smaller log\,$gf$ values were culled out. 
The results from the additions and screenings was that we could import 
some 2500 new potential line candidates to our atomic line library. 
With this step completed our library became suitable for synthesis of 
both O or B type spectra.  It is available upon request to the author.

\subsubsection{Line synthesis }

   All our line identifications were based on our just described 
line library. To summarize the result, all lines in our library have either 
measured or computed log\,$gf$ values taken from the above referenced 
sources. To make line identifications operationally we used a line annotation 
facility in the spectral line synthesis program {\sc SYNSPEC49} of Hubeny,
Lanz, and Jeffery (1994). This program can be run interactively to compute
and plot spectral fluxes over a specified wavelength range once the user
specifies key parameters such as metallicity, stellar effective temperature 
T$_{eff}$, log\,$g$, and microturbulence.  In our models we used 
log\,$g$ = 4, $\xi$ = 5 km\,s$^{-1}$, and solar abundances. 
We executed the manual steps outlined below three times to minimize errors.

We also compared line synthesis results from standard Kurucz (1990) and
non-LTE models calculated by TLUSTY described by Lanz \& Hubeny (2003),
as updated by a ``OGA grid" conveyed by T. Lanz to the author. The primary
improvement in the new grid was the addition of a number of excited levels
of light and iron group ions for non-LTE computations for lines of these ions.
The line strengths produced from LTE and non-LTE models typically differed
by amounts equivalent to a temperature change of 1,000\,K. 
The spectra computed from the LTE and non-LTE led to nearly identical
identifications. In those few cases where differences occurred, we took
the identifications from the non-LTE atmospheres models. Also, we erred 
on the side of reporting a possible identification rather than not doing so. 
In general the errors in the computed line strengths owe more to
to the low precision of the log\,$gf$'s than to uncertainties
in details of line formation in the atmospheres.

\subsubsection{Line identification methodology}
\label{linid}

  In this atlas we describe the philosophy and general methodology of
producing a list of line identifications predicted by model synthesis
programs. Those features observed in the template spectra at wavelengths
where no plausible identification can be made are annotated by the symbol
of a fictitious element, ``UN I" (for ``unknown"), in our tables. We
refer to these below as ``missed" features.
Without exception the photospheric spectra of our selected stars 
exhibit at least moderate broadening. Therefore, unidentified stellar 
features could be readily differentiated from the narrower atomic and 
molecular features.

In the far-UV a particular challenge is identifying components of 
closely spaced individual lines. 
To differentiate potential partially resolved lines we set 
a resolution window criterion of ${\pm 80}\,$m\AA.~ 
Lines within this 
window were considered for identification purposes to be completely blended.
Our procedure was to compute spectral line models in small wavelength intervals
and to overplot the synthesis and the identifications provided by
{\sc SYNSPEC} on a computer screen. Manual intervention was sometimes 
required in assembling the identification list for each template spectrum for
the following reasons:

\begin{enumerate}
\item The log\,$gf$ values of far-UV lines can be uncertain and occasionally
overpredict the line's observed strength; that is, a line is predicted by
{\sc SYNSPEC} but is not observed.  We introduced a ``:" symbol in the 
atlas figures and identification tables in cases where the overprediction 
exceeded a factor of 30 of the log\,$gf$ needed to make the predicted line
easily visible (i.e., a line depth of $\sim$10\%).

\item Closely spaced lines can blend in line syntheses. We chose a blend 
window of ${\pm 80}$ m\an]
to represent lines within a line
``group." In these groups we gave primacy to the dominant (primary)
line and listed the secondary lines (members of the common group) in a 
separate column of our identification table (see Tables\,3 and 4). 
In practice, almost all detectable far-UV lines are at least partially 
saturated. Thus, lines contributing less than the primary line's
absorption do not contribute much to a line group's aggregate strength.
Unless multiple secondary lines are present within a group,
we found that the strengths of ensemble group features are
dominated by the contribution from the primary lines.
\end{enumerate}

  The construction of our line list proceeded only after putting into place
semiautomated error checking procedures. One such procedure was to check 
the ions and exact wavelengths against those in our line library. Our 
program reported any errors in this collation, and they were corrected. 
Nonetheless, we cannot claim that our list of identifications is completely
error-free!  The influences of some lines may yet have been over-
or underestimated. Second, we checked our line identifications with
other published abbreviated lists of prominent far-UV lines, including
those in the Pellerin et al. (2002) atlas and the far-UV Capella
atlas (Young et al. 2001). We noticed minor deviations in quoted
wavelength values for several lines, and in a few cases we could
not confirm their line identifications because we could not find 
log\,$gf$ values in their secondary sources. Third, we checked the
major far-UV transitions noted for the light metal elements in the 
Grotrian diagrams published in the Bashkin \& Stoner (1975) compendium. 
The absence of any significant discrepancies in these comparisons suggests 
that there are few or no gross or systematic errors in our list.
Our experience in constructing two atlases suggests that the greatest
source of errors is in incorrectly estimating the relative strengths of 
photospheric and ISM (especially from atomic species) contributions.
Nonetheless, in all we believe our identified and unknown lines form 
a list of essentially all the visible features that contribute
the far-UV absorption lines of Galactic main sequence O4--O9.5 stars.

\section{The O star spectral atlas}
\subsection{The atlas and associated data products}

  Our O star atlas is different from most other atlases, though
it is similar in appearance to our B star atlas.
The atlas consists of three core products: (1) extensive line identification 
lists,
(2) a graphical plot of line identifications of the template spectra, and
(3) data files containing merged spectra in FITS format for all 25
atlas stars.  The FITS files
contain the Side 1 and Side 2 spectral data in two extensions. Each extension
contains a wavelength, flux, and pipeline-generated flux error vector. 

The full line identification tables and line-annotated spectral plots
for the six template stars, 
as well as the spectra in FITS format for all 25 stars (including those
drawn from the supergiant and Magellanic Cloud populations), have been placed, 
coincident with this paper's publication in MAST's High Level Science Products 
(HLSP) web area (http://archive.stsci.edu/prepds/fuvostars/), where the 
products are further vetted by MAST staff for clarity and ease of access. 
The full line lists and plots for the six template files are also accessible 
at this web site.  Users can report any errata, which MAST will include as 
part of the HLSP. 
In addition, a copy of our compiled line library will be provided upon 
request to the author.
Finally users of the HLSP or the individual {\it FUSE} archival data files will 
find electronic links to this paper held by the NASA Astrophysical Data System 
(ADS) and, conversely, links at the ADS site to the paper also include 
links to the {\it FUSE} data used herein.

Our detailed line lists for the O9.5--O7 and O6--O4 main sequence 
template spectra are presented as stubs in Tables\,3 and 4 in the paper
edition of this work.
The full tables are given in the electronic version of this paper in ASCII
Comma Separated Value format. 
Each subpanel lists the spectral type by star code, 1--6. The star code
listed applies only to identifications for the table's spectral subtype 
and later subtypes. 
In the next two columns we give the identified wavelength from our line list
catalogs and and finally the ion corresponding to the line identification.
The two wavelength columns represent either
the ``primary" or ``member" (secondary) of a line group, respectively, such
that one of the columns is always unfilled. Here ``group" refers once again to
lines occurring a common resolution window in which the primary line is located.
According to this definition, we found as many as 6 secondary group members 
associated with a primary line (generally no more than half of which are 
photospheric lines).
In these tables wavelengths of photospheric and ISM lines are 
given in the same (rest) system.

  The rows of Tables 3 and 4 are sorted according to the primary line 
wavelength of each group and are interleaved among the three template spectra.
The electronic edition of this paper gives this as a single monolithic 
table, with each spectral type subtable separated from the next by a header.
Secondary group members follow their associated group primaries, even if the
secondary's wavelength is slightly smaller than the primary's. This practice
insures that lines do not appear in more than in one column for a given star.

As in Paper\,1 we list in Table\,5 wavelengths of ``isolated" 
(least contaminated)
photospheric lines that are ``universal" (that is, visible in at least the
O5--O8 template spectra) for light elements and some iron-group ions.
In particular, note that the lines in these tables may lie further
away from H$_2$ features than our 
formal ${\pm 0.08}$\,\an\ resolution bin but still may be 
useless if the latter are pressure broadened.  Table\,6 gives isolated 
Fe\,V lines arising from the Fe$^{4+}$ ion, which as we will see is the dominant
contributor to photospheric lines for middle-O stars. The electronic edition 
of this paper presents this table in its full length, listing 135 lines.
The Fe$^{4+}$ ion continues to contribute lines in the 
far- and middle-UV to spectral types as late as B0.2 for main sequence stars. 
The tables also give the range of excitation potentials, in eV, of 
lines for an ion.  
The coding system in this table gives the same running star number (``1" for
O9.5, ``2" for O8, etc.) as for previous tables. A null (blank) value means 
that a line is not predicted in the spectrum. Blanks occur most commonly 
in the O4 and O9.5 star columns for low or high excitation ions, respectively.

\begin{figure*}[ht!]
\centerline{
\includegraphics[scale=.90,angle=90]{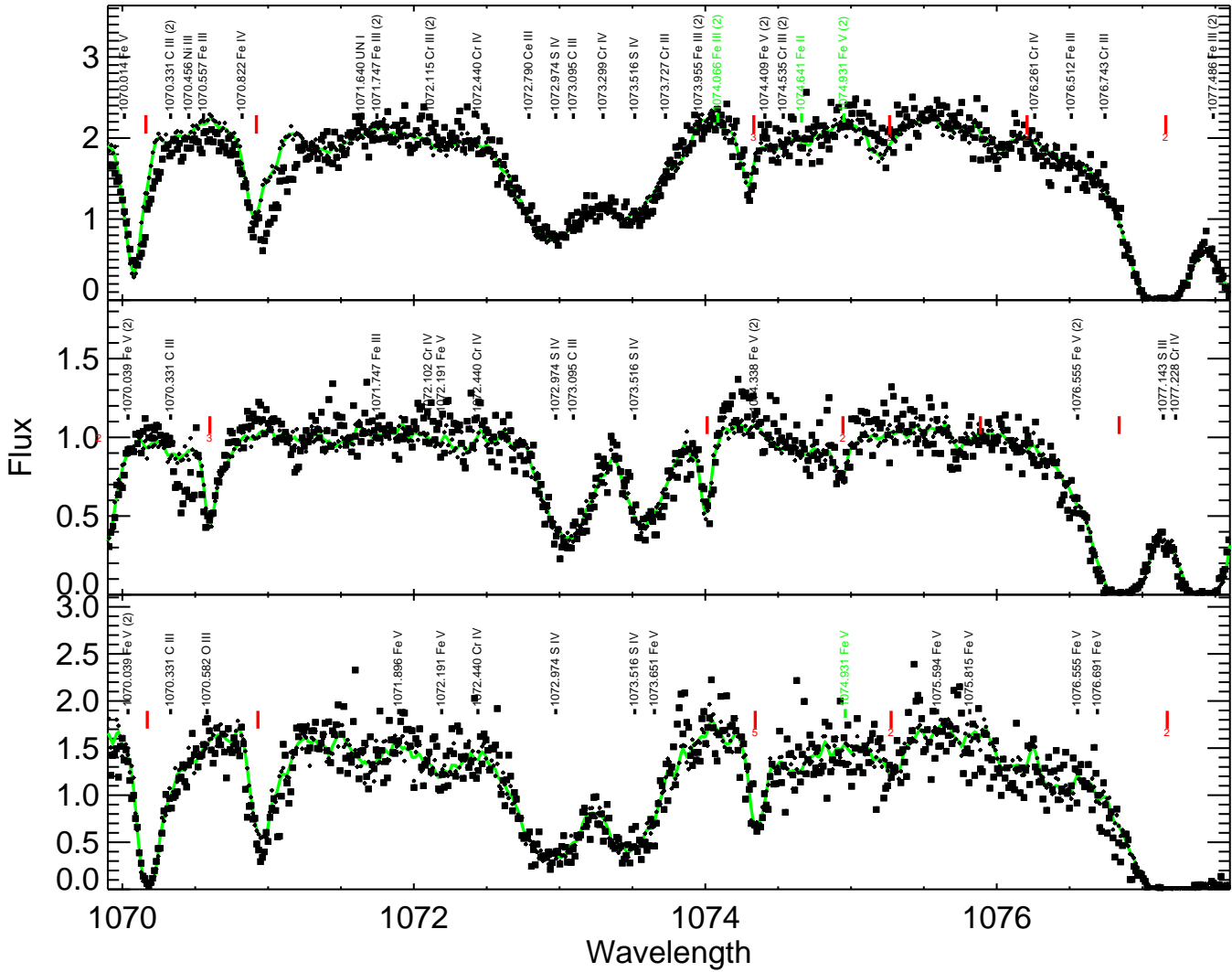}}
\vspace*{-.15in}
\caption{Spectral atlas and line identifications depicting the wavelength 
region 1070--1077.5\an\ for O9.5, O8, and O7 stars (top to bottom) near the 
main sequence. 
The green line represent a filtered {\it FUSE} spectrum taken from 
Side\,1 detectors; small squares represent Side 2 data. Line identifications are
represented vertically by ion and vacuum wavelength and with colons when they 
are uncertain.  Notations such as ``(2)" represent the combined number of 
primary and secondary lines in a wavelength resolution bin.
Annotations also represent atomic ISM lines while thick vertical 
ticks represent ISM H$_2$ features; the online plots show these features
in green and red, respectively. For the paper version wavelengths of 
atomic ISM lines in this and the following figure are followed by ``+ ISM."
Numbers are associated 
these too if they are primaries in local wavelength groups. 
Fluxes units are 10$^{-11}$ ergs s$^{-1}$\,cm$^{-2}$\,\AA$^{-1}$.
}
\label{plt1} 
\end{figure*}

\begin{figure*}[ht!]
\centerline{
  \includegraphics[scale=.9,angle=90]{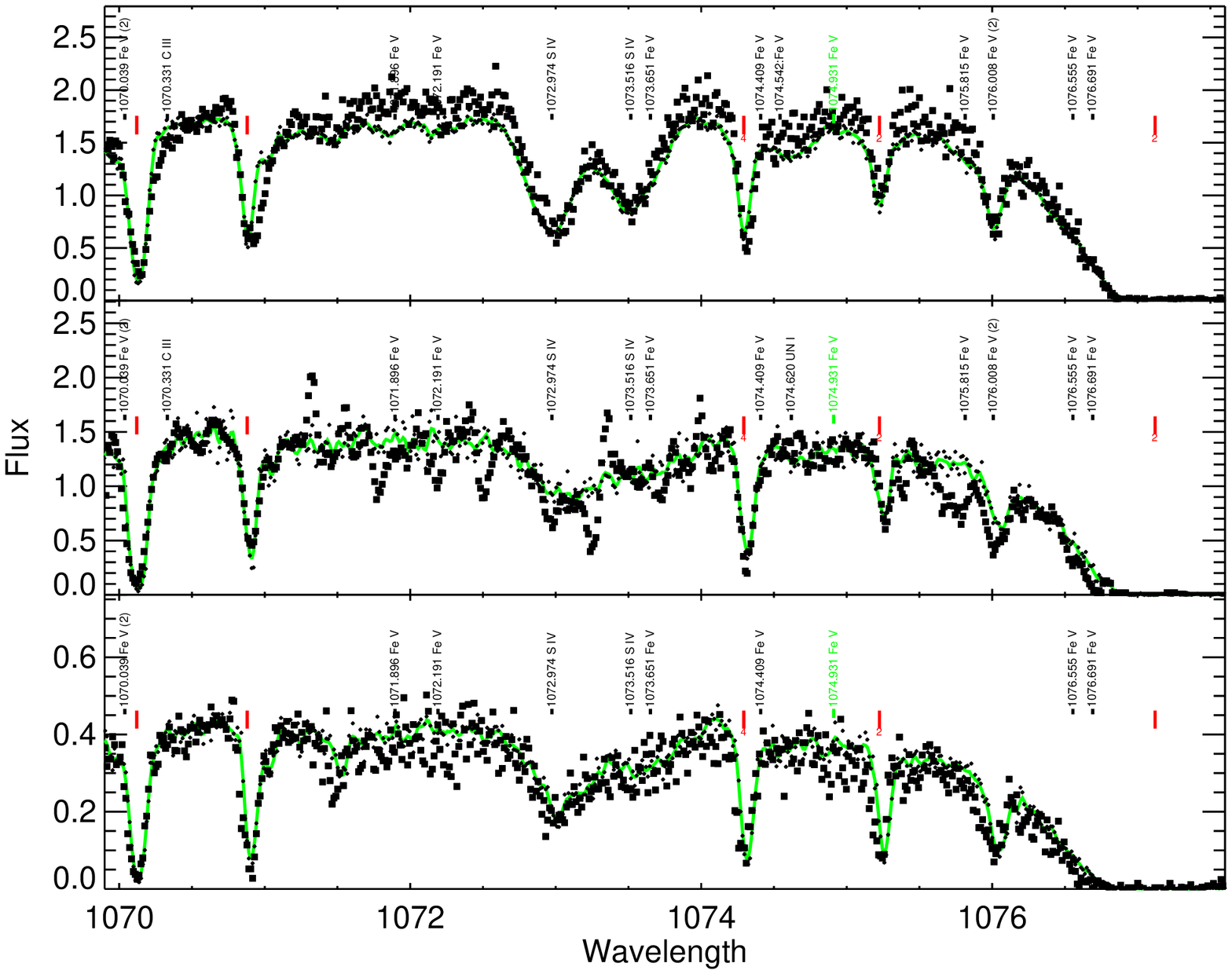}}
\vspace*{-.15in}
\caption{A continuation of the spectral atlas and line identifications in 
the wavelength region 1070--1077.5\an\ for O6, O5, and O4 stars (top to bottom) 
near the main sequence. Symbol identifications are the same as in 
Fig.\,\ref{plt1}.  The ion ``UN I" represents an unidentified absorption line 
not predicted by our atmospheres models.  
Fluxes are in units of 10$^{-11}$ ergs s$^{-1}$\,cm$^{-2}$\,\AA$^{-1}$,
except for the O4 star (bottom panel), for which the exponent is -12.
}
\label{plt2}   
\end{figure*}

Figures\,\ref{plt1} and \ref{plt2} depict 3-panel samples of the atlas 
over the 1070--1077.5\,\an\ wavelength interval for our O9.5, O8, \& O7 
and O6, O5, \& O4 template spectra, respectively.  This is the same
interval depicted in Fig.\,1 of the B star atlas and thus provides continuity 
for paper edition readers of the line coverage for O and B type stars.

\vspace*{0.30in}

\vspace*{0.3in}

  Because the stars have individual random radial velocities with respect 
to the local Galactic ISM system, we present in the figure the 
spectral lines and their annotations in their respective velocity systems, 
according to where the lines are formed. This is to say, in the figures
we correct for the stars' radial velocities such that the photospheric 
lines appear in our local rest frame whereas all H2 lines carry the reflex 
of the stellar spectrum's velocity. 
Two ramifications of this wavelength difference are, first, that
the composition of the mixed star-ISM line groups can differ for our
tables and figures. This is because the photospheric wavelengths in the 
figures are always represented in the laboratory system.  Second,
the ISM lines in the spectrum with the velocity shifts
will be displaced with respect to the positions of the same lines in
unshifted spectra shown in another figure panel. The best example of this
is the HD\,66788 spectrum (Fig.\,\ref{plt1}), for which as already noted 
the ISM lines are shifted relatively to photospheric lines by -0.6\,\an.~
Then, in the middle panel of Fig.\,\ref{plt1} 
the vertical ticks signifying the H$_2$ lines are displaced to the left by
this amount, whereas spectra in the upper and lower panels are unshifted.

\section{Discussion}
\subsection{Line statistics}
\begin{figure}
  \centering
  \includegraphics[,scale=.35,angle=90]{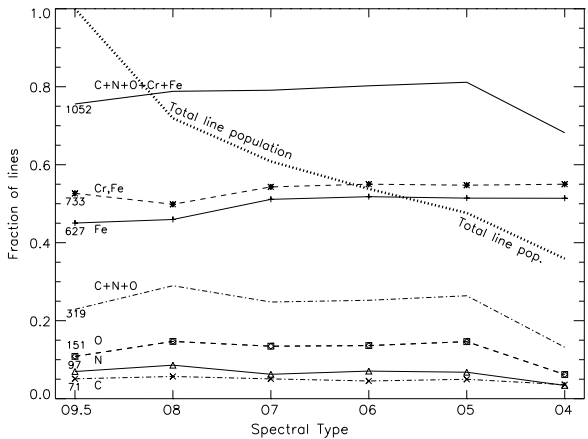}
\vspace*{0.2in}
\caption{
The run with spectral type of the fraction of identified far-UV photospheric 
lines for O stars near the main sequence. 
Fractions of lines for various elements are differentiated. 
Nearly 80\% of all lines arise from C, N, O, Cr, or Fe ions. 
Line and symbol types are used just to clarify these relations. 
The dotted line is the fraction of photospheric lines identified for 
each spectral type relative to the 1392 lines in the O9.5 spectrum.
Numbers under the O9.5-type fraction are the maximum {\it number} of 
lines for the element(s) for all O spectral types.} 
\label{frc}  
\end{figure}

Even a glance at the Tables 3 and 4 suggests that the density of 
photospheric lines, e.g., those in the range 949--960.0\an, 
decreases towards earlier types. For the same wavelength range
the number of lines for the B0.2 template was 81 (Paper\,1),
and in our tables it is 53 for O9.5\,V and 24 for O4\,V. 
Near the long wavelength limit, 1170-1188.5\,\an, the corresponding decline 
in numbers of identifications is as least as dramatic, from B0.2 to O9.5 to O4:
128, 97, and 31, respectively. Another important generalization is that with 
few exceptions the strongest features in the observed spectra are ISM or 
hydrogen Lyman lines -- that is, nearly all metallic lines are weak. One can
get a superficial impression that strong photospheric 
features appear in far-UV spectra of earlier type stars, 
but this is due to saturation of wings of strong H$_2$ 
features because of longer ISM column lengths to these distant stars. 
The numbers of all identified lines in our Tables\,3 and 4, including ISM ones, 
total 1798, 1359, 1178, 1077, 992, and 821, for O9.5, O8, O7, O6, O5, and O4,
respectively.
When we separate out the ISM lines the totals for 
photospheric lines alone are 1392, 1001, 847, 749, 663, and 500,
respectively. The totals are depicted by a thick dotted line in 
Figure\,\ref{frc}.
As one proceeds to earlier O spectral types, the downward trend shown by
these totals contrasts with the nonmonotonic numbers of lines found along 
the B spectral sequence in Paper 1.  In the comparable 
far-UV wavelength region those numbers were 2182 (B8), 1398 (B2), 
and 1991 (B0.2). Not surprisingly, one finds that there is a fair degree 
of overlap among lines of neighboring spectral subtypes. 
The total number of unique photospheric identifications is 1792. The number 
of photospheric lines we identified for all six O spectral types is 300. 
Note that because our stellar line 
identifications are based on line synthesis predictions (except for 
missed features seen in the spectra), our O star statistics are not 
affected by the increased line broadening in our template spectra with 
respect to the (generally smaller) line broadening encountered for our 
B star spectra presented in Paper\,1.

  The fundamental cause of the changes in these line totals is the shifting 
ionization states with increased effective temperature. 
This is easiest to see for lines of the iron group elements because
their differences in ionization potential are relatively modest for 
successive ion states.
As we noted in Paper\,1, the changes in B star spectra occur 
because the total number of visible Fe\,II plus Fe\,III lines in the B2 
far-UV spectrum is lower than the totals of Fe\,II plus Fe\,III lines 
for either B0.5 or B8 spectra. 
Thus, early and late type B spectra exhibit more lines that are 
mainly either Fe\,II lines or Fe\,III lines, respectively, 
than the sum of the roughly equal numbers of 
Fe\,II and Fe\,III lines in a B2 spectrum.
This type of ``rule" also holds true for the rapidly changing Fe ionization
states among types O9.5 through O6. 
In fact, the O9.5 spectrum exhibits a total of 656 Fe lines, 
about half (324) of which are Fe\,III lines.
However, as we progress to O8, O7, and O6 types, Fe$^{4+}$ has become
the dominant state and 60-65\% of the visible iron lines are Fe\,V.  
By contrast, when we reach subtypes O5 and O4, even though the 
dominant ion stage has shifted from Fe$^{4+}$ to Fe$^{5+}$, 
most visible iron lines in the far-UV are still Fe\,V lines.  
The ultimate cause of these changes in our statistics in the far-UV 
is that the visible Fe\,VI and Fe\,VII lines 
have moved into the extreme UV as the typical energy differences between
transition energy levels increases with ionization potential. Extreme-UV lines
are no longer visible to us, owing to the high ISM opacity at Lyman continuum 
wavelengths. Also, the dominant carbon and nitrogen 
(and to a lesser extent oxygen) atoms have been stripped of all or nearly all 
of their electrons and contribute relatively few lines at far-UV wavelengths.

  Another attribute of our line statistics is the similar percentages
of secondary photospheric lines relative to the total membership of a
line group remains about the same:~ 61\% for O8--O9.5 and 64-66\% for O4--O6. 
In B stars (Paper\,1) the percentages are roughly the same, 
59--66\%.\footnote{We omit the $\tau$\,Sco identifications from this 
statement because in Paper\,1 we used a narrower wavelength window, 
${\pm 0.05}$\,\an,,~ for assignment of lines to co-blended line ``groups."}
This means that the degree of blending of photospheric lines within 
predominantly photospheric line ``groups" is similar for both B and O
spectra in the far-UV.

We were pleased to discover that the percentage of unknown (i.e., ``UN~I," 
or underpredicted) lines found relative to successfully predicted 
photospheric lines decreased in O star 
spectra relative to the B star spectra in Paper\,I. These percentages
come to 1\% for O7--O9.5 stars, jumping only to 3\% for O4--O6.
On the other hand, the fractions of overpredicted (predicted but not 
observed) lines in the B star lines were high (6--12\%). We find that this
population declines from 3\% at O9.5 to a minimum of near 1\% for O6--O8
stars and increases again to more than 3\% for O4--O5.
These low percentages are all the more surprising when one recalls 
that the ``overpredicteds" are features we have {\it observed} in the spectrum, 
whereas our ``identified" features are {\it predicted} from 
synthesis models -- whether they are visible in spectra or not. These
percentages also speak well to the relative completeness of the published 
libaries for far-UV lines arising from multiply ionized ion states.

\begin{figure}[ht!]
\centerline{
  \includegraphics[scale=.35,angle=90]{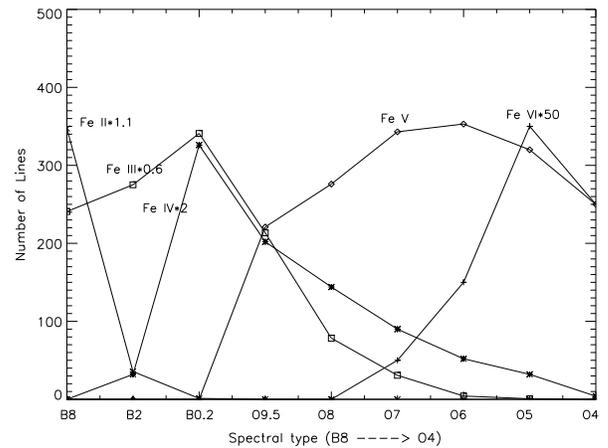}}
\vspace*{0.2in}
\caption{
The run of Fe\,III through Fe\,VI lines with spectral type for
identified photospheric lines in the B star and O star (this paper) atlases.
For ease of reference, each line is scaled to the same maximum as the 
Fe\,V population. The numbers following the * symbol are these scaling factors.
} 
\label{fenums} 
\end{figure}

 The general trend of line populations arising from the most prolific
line-producing elements in the far-UV spectrum, namely Fe, Cr, C, N, and
O, is exhibited in Fig.\,\ref{frc}.
Notice from this plot that 75--80\% photospheric lines in the far-UV
are caused by these five elements. The fractional contributions from these 
elements to the total number of lines remains nearly the same with spectral 
type. The exceptions are the numbers of N and O features: the numbers of 
N\,III and O\,III lines drop off at O5--O4 because of shifts in ionization.
By comparison, the numbers of N\,IV and O\,IV--O\,VI 
lines in spectra of these hotter stars are negligible.

  Fig.\,\ref{frc} also shows that iron ions alone contribute at least half the
visible photospheric lines in far-UV spectra of O stars. This is somewhat 
higher than the corresponding fractions (34\%--47\%) in the B star 
atlas.  Table\,5 of that atlas shows the relative numbers of lines from 
various iron ions. As already mentioned, we found 
that the numbers of Fe\,III (and to a lesser extent Fe\,IV) lines 
peak at B0.2. Similarly, the Fe\,II lines were found to peak at B8. 
In Figure\,\ref{fenums} we present combined totals for all iron ions 
that contribute to the far-UV spectra of B stars and O stars. 

This plot shows that whereas the
``big story" for B star spectra was the transition from Fe\,II to Fe\,III
lines, for O star spectra it is the dominance of Fe\,V lines. This occurs
once the transition from Fe\,III to Fe\,V lines sets in,  at subtype O9.5.
Fe\,II lines are nonexistent in the photospheric spectra of O stars.

\subsection{Temperature and chemical Indicators}
\label{chem}

  We end our description of this atlas by discussing 
specific far-UV lines that can be used in O star main sequence spectra 
to define new diagnostics of effective temperature and in some instances 
chemical composition. For O stars many of the most interesting 
abundance anomalies arise from interior CNO-cycle processes.
However, iron group abundance differences
could also be affected by reprocessing in recent supernova explosions. 
Therefore, we also explore the isolated lines already introduced in 
Tables\,5 and 6 for a number of ions. 
For the star codes of these tables, decimal values ending in ``.1," 
such as ``1.1," signify that a dominant line is closely blended with 
a line of the same ion, i.e., the blend is effectively one strong line. 
We refer to these as ``self-blended" features.

\noindent {\it Diagnostic lines:}

\noindent {\bf He\,I:~}  All five He\,I lines in Table 5 are visible in all 
O template spectra, but most of them are severely blended with local and
often broad H$_2$ features. The best T$_{\rm eff}$ indicators are the
lines at 958.7\,\an\ and 1084.9\,\an. The strengths of both peak at O6--O7. \\

\noindent {\bf C\,II:~}  We found only one completely isolated C\,II line, 
1010.371\,\an,~ in our spectra of O9.5--O6\,V dwarfs. A weaker member of this
6\,eV multiplet at 1010.083\,\an\ can sometimes be discerned in the red 
wing of a strong H$_2$ feature at 1010\,\an\ in redshifted spectra. 
Two other features at 1065\,\an\ are present in narrow-lined spectra of
late O stars, but they are overwhelmed by local H$_2$ features.  
We do not recommend 
attempting to use C\,II lines for quantitative analyses of O star spectra.  

\noindent {\bf C\,III:~}  The C\,III lines peak at subtype O9.5 in our
main sequence template spectra. A single line at 1070.331\,\an\ is optically
thin but lies within 0.3\,\an\ of a H$_2$ feature. Therefore it cannot
be used for studies of carbon abundances, at least for stars having small
Doppler shifts. With the exception of 977.020\,\an,~ the C\,III 
lines shown in Table\,5 (1125\,\an--1176\,\an) are blended with other
weak multiplet members. 
As we pointed out in Paper\,1, the
wings of the 977\,\an\ line are sensitive to electron pressure for all
O spectral subtypes and thus can be used with the C\,III 1175.9--1176.3\,\an\ 
multiplet lines to determine a star's luminosity class.  According to 
Table\,5 and Figure\,\ref{plt1}, the wings of the  1174--1176\,\an\ 
complex can be blended by Fe\,IV and Fe\,V lines at types O7--O9.5. Thus,
equivalent width ratios measured from these lines and 977\,\an\ are best 
formed by including only the cores of the broad 1174-1176\,\an\ features.
Finally, as already noted, in supergiant spectra the C\,III complex 
has transitioned to a P\,Cygni profile, 
and the photosphere no longer makes a significant contribution.

\noindent {\bf C\,IV:~} The most visible lines of C\,IV lie close together 
at 1168.847\,\an\ and 1168.999\,\an\ and so may often be treated  as a 
single feature. Thus, the strength of the feature peaks at subtypes O6--O7.
In Paper\,1 we noted that the 1168\,\an\ line makes its appearance at subtype 
B0.2, and along with the nearby C\,III complex can form a good diagnostic
if the temperature is isolated by lines of other ions. This is also true
in the late O stars for classes III-V. As one proceeds to hotter temperatures
the C\,IV line becomes considerably stronger, making it more sensitive
to temperature than luminosity. A much weaker isolated feature lies at 
1097.319\,\an\ and begins to fade at O5.

\noindent {\bf N\,III}: Two isolated N\,III lines are visible at
1103.044\,\an\ and 1106.036\,\an, but they are very weak and therefore are not 
good diagnostics.  Both 1112.648\,\an\ and 1152.406\,\an\ lines lie close 
to ISM features and therefore can be used at best with caution. 
The best N\,III diagnostics for temperature or abundance in the far-UV
are 1183.032\,\an\ and 1184.514\,\an.~ However, they are self-blended 
by other N\,III lines with similar log\,$gf$ values.  

\noindent {\bf N\,IV:~} Two isolated N\,IV lines are visible at 1131.488\,\an\ 
and 1133.121\,\an.~ An advantage provided by these lines is that their 
strengths do not vary significantly with spectral type. Therefore the lines 
can serve as nitrogren abundance indicators for all luminosity classes.

\noindent {\bf O\,III:~} Lines of this ion are numerous in the optical-UV
spectra of B stars, but they nearly disappear in O star spectra.
Nonetheless, the oxygen and iron group abundances in these stars 
tend to track well, and a few remaining O\,III lines can be
used, especially with Fe\,V lines, for temperature determinations. 
The combined strengths of three isolated O\,III lines at
1007--1008\,\an\ in practice renders them a single feature even in 
moderately broadened (v$sin\,i$ $\sim$ 100\,km\,s$^{-1}$) O type spectra. 
The best lines for diagnostics are those that are self-blended, 
such as 1149.602\,\an\, 1150.882\,\an\, 1153.022\,\an\, 1160.154\,\an\, and 
1172.451\,\an,~ and two isolated lines at 1157.041\,\an\ and 1157.161\,\an.~
However, even the latter lines generally overlap due to the velocity broadening.

\noindent {\bf O\,IV:~} Although less sensitive to temperature than O\,III 
lines, the excitation potentials $\chi$$_{exc}$ of O\,IV lines are in
the range 48--63\,eV. The lines increase in strength as one goes to earlier 
types.  The lines at 1045.364\,\an,~ 1046.313\,\an\,~ and 1080.969\,\an\ 
are isolated but weak. The lines 1067.832\,\an,~ (self-blended) and 
1067.959\,\an\ are otherwise isolated, but they may blend from velocity 
broadening.  
The 1080.969\,\an,~ 1164.546\,\an,~ 
and 1167.478\,\an\ lines are good diagnostics too, but they arise from higher
excitations and disappear at subtype O4.  In terms of strength and isolation, 
the best features among these candidates are the two 1067--1068\,\an\ lines.

\noindent {\bf O\,V:~} 
Three O\,V lines given in Table\,5, 995.087\,\an,~ 1010.602\,\an,~ and
,1090.320\,\an\ all appear weak in our O main sequence spectra, but otherwise
they are viable temperature diagnostics for most of our template spectra.
Because the ionization fraction decreases with advanced spectral type, the
O\,V lines become relatively weak at O9.5\,V. At this subtype 995.087\,\an\ 
becomes dominated by neighboring Fe\,III lines and cannot be used. We remind
the reader that the typical $\chi$$_{exc}$'s for most O\,III lines
in our sample lie in the range 24--36\,eV. The $\chi$$_{exc}$ is
single-valued at 92\,eV for O\,V lines. 
As long as O\,III and O\,V lines can be found in an O star spectrum, 
their ratios can provide good measures of the ionization in the photosphere.  

\noindent {\bf O\,VI:~} The 1031.926\,\an\ and 1037.617\,\an\ resonance 
doublet lines have been known as critical diagnostics of mass loss in O stars 
and B supergiants since their discovery from UV rocket experiments long ago. 
Their anomalous strengths are due to the ``superionization" of 
this ion, a phenomemon which has long been ascribed to the atomic Auger 
effect from X-ray irradiation. However, it now appears instead that the
strengths may be due to clumping of
parcels within a tenuous component of the O star winds (Zsarg\'o et al. 2008).
The doublet is badly marred by local H$_2$ and nearby Fe\,V lines, and 
for this reason we do not list them in Table\,5 as isolated lines for main 
sequence stars. Their strengths grow rapidly with temperature and luminosity, 
as O$^{5+}$ becomes the dominant ionization state. 
Save for late O main sequence spectra, the wind contribution dominates the
formation of the resonance lines across most of the profiles. Their wings 
are typically in emission and asymmetric.

\noindent {\bf Si\,III:~} Visible lines of both Si$^{2+}$ and Si$^{3+}$
have the advantage of being strong although they are nearly all self-blended. 
For practical purposes the sole representative Si\,III line is 1108.358\,\an.~

\noindent {\bf Si\,IV:~} Except for minor contaminations by local
weak Fe features, the Si\,IV
1066.614\,\an\,, 1122.485\,\an,~ and 1154.621\,\an\ lines can be used along 
with Si\,III as temperature diagnostics. Since the $\chi$$_{exc}$'s for 
these lines are $\le$31\,eV, these lines form a low-excitation complement to 
the more excited oxygen lines as diagnostics of photospheric temperature.

\noindent {\bf P\,IV:~} To varying degrees all the P\,IV lines in Table\,5 
are isolated in our O spectra, even given moderate velocity broadening.
The strengths of all these lines grow with P$^{3+}$ ionization fraction
and atmospheric temperature.
With two caveats the best lines for temperature diagnostics in main 
sequence O star spectra are 1118.552\,\an\ and 1187.540\,\an.~ The first 
caveat is that the 1118\,\an\ line is one of the lesser isolated lines in 
this small group and is often overwhelmed by a strong P\,V neighbor. 
The second caveat is that the
instrumental sensitivity of the single LiF Side\,1B {\it FUSE} detector 
is low where the 1187.540\,\an\ line is recorded.  
We point out that the 1187\,\an\ line can often be seen in well exposed 
{\it IUE} and {\it HST} spectra of O stars. 
Also, from our line synthesis we believe the line displayed at
the position of 1187.5\,\an\ in the Brandt et al. (1998) atlas of the 
O9\,V star 10\,Lac is actually this P\,IV resonance line and not an
Fe\,V line they identified.

We also note that the atmospheric contribution to the P\,IV resonance 
line at 950.657\,\an\ is stronger than the ISM component, at least in most
of our template spectra. In our spectral atlas identifications 
for O7--O9.5 stars the ISM component 
of this same transition is blueshifted to an apparent position of 
$\approx$1150.45\,\an.~ We can see there that the ISM component is weaker 
than the same line in the photosphere of the O8 template spectrum (see
continuation of Figure 1 in the on-line edition).  
This is very likely true for the O9.5 and O7 spectra as well.

\noindent {\bf P\,V:~} 
As we know from the Pellerin et al. (2002) O star atlas, the
P\,IV 1118.043\,\an\ and 1128.010\,\an\ are often among the strongest
features in the far-UV spectra of O stars. The wind contribution is especially
important. For example, 
in their detailed fittings of the profiles of these lines, Fullerton, Massa, 
\& Prinja (2006) noted that near O7 the photospheric component even near the 
line core is generally dominated by the wind component except in those cases 
where the wind is weak.  This fortunate circumstance occurs because of the
low abundance of this low-alpha element.
Substantial P\,Cygni emission can be seen in the 
wings of supergiant spectra among all subtypes O4--O9.5.  
Both resonance lines are strong, and yet they are usually optically thin 
in the winds of luminous O stars. These conditions make them superb 
diagnostics of mass loss from radiative winds.  
Fullerton et al. also found that a nearby Si\,IV 1128 line contributed to 
the red wing of this doublet substantially in O7 spectra for all 
luminosity classes.  It should be noted that a weak P\,V line at 
1000.358\,\an\ is also weakly present in all our template spectra.

\noindent {\bf S\,IV:~} The most important S\,IV lines are the resonance
lines at 1062.662\,\an,~ 1072.974\,\an,~ and 1073.516, which are exhibited
in the Pellerin et al. atlas. All of these attain maximum strengths at about 
O7.  Pellerin et al. also noted the presence of S\,IV lines
at 1098.362\,\an,~ 1098.929\,\an\,, 1099.482\,\an,~ and 1100.051\,\an.~ 
These lines are visible up to O6 in our main sequence templates, even though
nearby H$_2$ contributions often overwhelm the 1099\,\an\ and 1100\,\an\ lines.
``Also rans" as temperature diagnostics are the weak 1106.487\,\an\ 
line and the  stronger 1110.905\,\an\ and 1117.161\,\an\ lines.
Except for the resonance lines, 1138.076\,\an\ and 1138.210\,\an\ 
are the strongest isolated S\,IV lines.  
interior nucleosynthetic processes, the S\,IV lines are good diagnostics 
of temperature.

\noindent {\bf Fe\,III:~} The only isolated Fe\,III line surviving among
the O stars (to O6) is 1091.082\,\an.~ However, it is a weak feature.
Even at O9.5 it merges with a nearby P\,IV line in most of our spectra. 
The Fe\,III lines are of little help as physical diagnostics in O star
spectra.

\noindent {\bf Fe\,IV:~} The dominances of odd iron ion stages, 
such as Fe$^{3+}$ are narrowly peaked in temperature. 
As Fig.\,\ref{fenums} suggests, the strengths of Fe\,IV lines 
peak at subtype B0.2 in main sequence spectra and decrease
into the middle-O star spectra.  Nonetheless, several far-UV lines 
can be utilized as temperature diagnostics. The most useful ones
are the isolated, albeit weak, lines at 1005.697\,\an\ and 
1156.526\,\an.~ The strongest Fe\,IV feature is 1170.781\,\an.~ 
All Fe\,IV lines disappear by O6 or O5.

\noindent {\bf Fe\,V:~} This ion contributes by far the largest number 
of lines to far-UV O star spectra (Fig.\,\ref{fenums}). 
In part because there are so many Fe\,V lines, few individual 
features are strong  - even in middle-O stars where the Fe$^{4+}$ 
ionization fraction peaks. However, one strong unblended line is 
1007.292\,\an.~  For this case even nearby lines are comparatively 
weak. Two moderate strength Fe\,V lines that may provide useful
estimates of the iron abundance 
are 1011.367\,\an\ and 1011.512\,\an,~ but these merge in broad lined spectra. 
The same is true of the Fe\,V lines at 1018.059\,\an\ and 1018.198\,\an.~
Being nearly isolated and of moderate strength, 1043.991\,\an\ is another
potentially useful line.
The shortest wavelength isolated Fe\,V lines are 958.288\,\an\ and 
958.379\,\an.~ These lines are weak but visible across 
the O4-O9.5 domain. These lines should be used with caution because
of the nearby He\,I line.

\noindent {\bf Fe\,VI:~} This ion represents the most excited ionization 
state among all iron group elements in our survey. 
A few isolated lines are visible in O4 to O6 spectra:
1160.561\,\an,~ 1170.275\,\an,~ and 1186.575\,\an.~ As with Fe$^{3+}$, 
the dominance of Fe$^{5+}$ extends over a relatively small temperature 
range.  The strengths of these lines begin to weaken in O4 spectra. 
This trend continues in spectra of the hottest (O2 and O3) stars.

We hope this atlas will serve its intended purposes to aid in the  
analysis of O star spectra, including population synthesis of star
clusters containing massive stars.
The author may be contacted for additional information 
either directly or via MAST.

\acknowledgments
We would like to express our thanks to Dr. Thierry Lanz for his assistance
in the compiling and running of {\sc SYNSPEC49} for non-LTE O-star atmospheres
models on the author's computer and Dr. Alex Fullerton for his advice on
O star physics and encouragement. Suggestions by the referee markedly
improved the quality of the manuscript.
This work was supported by a NASA Astrophysics Data Analysis grant NNX10AE58G 
to the Catholic University of America.

\clearpage

\begin{table*}[ht!]
\tablenum{1}
\begin{center}
\caption{\centerline{Star list for O star LUV spectral atlas}}
\label{strr}
\begin{tabular}{lrrrr|lrrrr}
\tableline\tableline
Star          & Sp. Type    & Ref   & $vsini$ & $vsini$Ref   &   Star     &  Sp. Type &  Ref  & $vsini$ & $vsini$Ref   \\
\tableline 
O2            &    &   &  &   &  O7  &   &   &  &    \\
BI237         & O2V(($f*$)) & M05   &   83  & PG09  & HD 93222$^*$ &   O7 III ((f))  &   M04 &  60   &  H97    \\
MPG355        & O2III($f*$) & W02   &  112  & PG09  & Sk\,80       & O7 Iaf &  R03 &  76   &  PG09   \\
              &             &       &       &         & AV\,207    &   O7 V    &   J01 &  75   &  PG09   \\
              &             &       &       &         &            &           &       &       &         \\
O3            &             &       &       &         &   O8       &           &       &       &         \\
HD 93250      & O3V((f))    &  W72  &   107 &  PG09   &  HD\,66788$^*$ &    O8 V   &   R03 &  60   &  PG09   \\
Sk\,-68\,137    & O3III($f*$) &   W77 &   107 &  PG09   &  AV\,469$^2$   &    O8 II  &   G87 &  79   &  PG09   \\
              &             &       &       &         &  D301-NW8  &    O8 V   &   M95 &  57   &  PG09   \\
              &             &       &       &         &            &           &       &       &         \\
O4            &             &       &       &         & 09         &           &       &       &         \\
HD\,46223$^*$  &  O4 V((f))  & M73   &   82  &  PG09   & HD 46202   &  O9 V     &   W73 &  37   &  H97    \\
Sk\,-67\,166   & O4 I f$^{+}$ & S99  &    97 &  PG09   & HD 207198  & O9 Ib-II  &   W76 &  91   &  H97    \\
AV\,435       &    O4 V     &   F88 &    96 &   PG09  &     AV378  &  O9 III   & W02   &  64   &  PG09   \\
              &            &        &       &         &            &           &       &       &         \\
O5            &            &        &       &         &  O9.5      &           &       &       &         \\
HD\,46150$^*$ & O5 V(($f$)) & W73 &  111 & PG09  & HD\,112784$^*$ & O9.5 III & G77   &  51   &   H97   \\
HD\,93843     & O5 II       & G77   &   95  &   PG09  & HD\,152405 & O9.7 Ib-II   &   W76   &    77    &    H97  \\
MPG342        & O5-O6       &  W02  &  58   & PG09    & AV\,238    & O9.5 II   & W02   &  71   &   PG09  \\
              &             &       &       &         &            &           &       &       &         \\
O6            &             &       &       &         &            &           &       &       &         \\
HD\,42088$^*$ &  O6.5 V     & W72   &  65   &  PG09   &            &           &       &       &         \\
AV\,377       &  O6 V      & G87   &  51   & PG09    &            &           &       &       &         \\
Sk\,-66\,100    &  O6 II(f)   & W02   &  73   & PG09    &            &           &       &       &         \\
              &            &       &        &         &            &           &       &       &         \\
\tableline
\end{tabular}
\begin{list}{}{}

\item []{\it Note ``*":~~}Galactic main sequence stars used for
line identification lists.

\item []{\it Note 1:~~}Spectral Type and $vsin\,i$ reference codes are given 
in the bibliography.

\item []{\it Note 2:~~} MPG (NGC346), AV, and Sk stars are located in the Magellanic Clouds. D301-NW8 is also named LMC2-755.
\end{list}
\end{center}
\end{table*}

\begin{table}[ht!]  
\tablenum{2}
\begin{center}
\center{\caption{Wavelength coverage of FUSE detectors (\AA ngstroms)}}
\begin{tabular}{cc|cc}
\tableline\tableline
Segment  & Side 1    & Segment  & Side 2 \\
\tableline
SiC 1B & {\em 907-992} & SiC2A & {\em 917-1007}   \\
     & (930-990.45)    &       & (930-1005) \\
LiF 1A & {\em 988-1082} & LiF2B & {\em 984-1072}   \\
     & (990.45-1082)    &       & (1005-1071.35) \\
SiC 1A & {\em 1004-1092} & SiC2B & {\em 1016-1103}   \\
     & (1082-1090)    &       & (1071.35-1087.5) \\
LiF 1B & {\em 1074-1188} & LiF2A & {\em 1016-1103}   \\
     & (1094.5-1188)    &       & (1087.5-1181, \\
     &                   &       &  1090-1094.5$^*$)    \\
\tableline
\end{tabular}
\end{center}
\tablecomments{$^*$Side 2 data for the interval 1090-1094.5\,\AA~ are
used again for the Side 1 spectrum.}
\end{table}

\begin{table*}[ht!]  
\tablenum{3}
\begin{center}
\center{\caption{Line identifications in O9.5, O8, and O7 spectra}}
\begin{tabular}{rrrl|rrrl|rrrl}
\tableline\tableline
(a)  &  &  & (b) &  &   & (c)  &   &   & \\
In & O9.5 prim. & Sec. & Ion &
In & O8 prim.   & Sec. & Ion &
In & O7 prim.   & Sec. & Ion \\
Star \#1 & wave. & wave & Ident &
Star \#2 & wave. & wave & Ident &
Star \#3 & wave. & wave & Ident \\
\tableline
123& 948.917 & & Fe\,III   &    &       & &            &   &         & & \\
 &        & &          &  23&948.965& & Ar\,V      &23 & 948.965 & & Ar\,V \\
 &         & &           & 123&& 948.917& Fe\.III &  123 & & 948.917 & Fe\,III\\
1& 949.078 & & Fe\,III   &     & &         &         &   &         & & \\
 &         & &           &   23&  949.125 & & N\,III &23 & 949.125 & & N\,III \\
13& 949.181 & & H2\,I ISM &     &          & &      &13&949.181&&H2\,I ISM \\
1&&949.236 & Mn\,III     &     &          & &        & & &       & \\ 
123&949.351 && H2\,I ISM & 123& 949.351 & &H2\,I ISM &123&949.351& &H2\,I ISM \\
1&&949.328 & He\,II      &  2 & & 949.328 & He\,II  & 3 & &  949.328 & He\,II \\
 &         & &           &  23 & & 949.379 & Ar\,V  & 23 & &  949.379 & Ar\,V \\
123&949.603&&H2\,I ISM & 123 &949.603&& H2\,I ISM & 123 & 949.603&& H2\,I ISM \\
123&949.743 && H\,I ISM & 123& 949.743 & & H\,I  &123 & 949.743 & & H\,I \\
123&950.072&&H2\,I ISM & 123 & 950.072& & H2\,I ISM &123& 950.072& & H2\,I ISM\\
123&950.314&&H2\,I ISM & 123& 950.314 && H2\,I ISM&123 & 950.314 &&H2\,I ISM \\
123&&950.337&Fe\,III   & 123 & &950.337 & Fe\,III &123 & &950.337 & Fe\,III \\ 
123&950.397&& H2\,I ISM &123 &950.397 & & H2\,I ISM &123 &950.397 && H2\,I ISM\\
123&950.657 & & P\,IV  & 123 &950.657 & & P\,IV    &123 & 950.657 & & P\,IV \\
123&950.816 &&H2\,I ISM &123 &950.816 & & H2\,I ISM &123&950.816 & & H2\,I ISM\\
 &        & &            & 2 & 950.925 &     & Fe\,V & &         & &         \\
 &        & &            & 2 &    & 950.951 & Fe   V & &         & &         \\
\tableline
\end{tabular}
\end{center}
\tablecomments{Table\,3 is presented in its
entirety in the MAST archives at http://archive.stsci.edu/prepds/fuvostars/.
A portion is shown here for guidance in data format and content.}
\end{table*}

\begin{table*}[ht!]  
\tablenum{4}
\begin{center}
\center{\caption{Line identifications in O6, O5, and O4 spectra}}
\begin{tabular}{rrrl|rrrl|rrrl}
\tableline\tableline
(a)  &  &  & (b) &  &   & (c)  &   &   & \\
In & O6 prim. & Sec. & Ion &
In & O5 prim.   & Sec. & Ion &
In & O4 prim.   & Sec. & Ion \\
Star \#4 & wave. & wave & Ident &
Star \#5 & wave. & wave & Ident &
Star \#6 & wave. & wave & Ident \\
\tableline
456& 948.965 & & Ar\,V & 456 & 948.965 & & Ar\,V &456 & 948.965 & & Ar\,V \\
4& &948.917 & Fe\,III &  &         & &       &  &          & &        \\  
456&949.351 & & H2\,I ISM &456 & 949.351 &&H2\,I ISM &456& 949.351&&H2\,I ISM \\
456&&949.328 & He\,II  & 456 & & 949.328 & He\,II& 456 & &  949.328 & He\,II \\
456&&949.379 & Ar\,V  & 456 & & 949.379 & Ar\,V & 456 & & 949.379 & Ar\,V \\
456&949.603&&H2\,I ISM& 456 & 949.603& & H2\,I ISM &456 &949.603& & H2\,I ISM \\
456&949.743 && H\,I ISM & 456 & 949.743 & & H\,I  &456 & 949.743 & & H\,I \\
456&950.072 && H2\,I ISM &456& 950.072& & H2\,I ISM&456& 950.072& & H2\,I ISM\\
456&950.314 && H2\,I ISM&456 & 950.314 && H2\,I ISM&456 & 950.314 &&H2\,I ISM \\
4&&950.337 & Fe\,III     &   & &        &            &  & &        &         \\ 
456&950.397&& H2\,I ISM&456 &950.397 & &H2\,I ISM &456 &950.397 & & H2\,I ISM\\
456&950.657 & & P\,IV   & 456 &950.657 & & P\,IV    &456 & 950.657 & & P\,IV \\
456&950.816 & & H2\,I ISM &456&950.816 & & H2\,I ISM &456&950.816 && H2\,I ISM\\
456&951.073 &&Fe\,V &456 &  951.073 & & Fe\,V & 456 & 951.073 & & Fe\,V  \\
456&951.294 &&Fe\,V &456 &  951.294 & & Fe\,V & 456 &  951.294 & & Fe\,V \\
456&951.449 &&Fe\,V &456 &  951.449 & & Fe\,V & 456 &  951.449 & & Fe\,V \\
\tableline
\end{tabular}
\end{center}
\tablecomments{Table\,4 is presented in its
entirety in the http://archive.stsci.edu/prepds/fuvostars/.
A portion is shown here for guidance in data format and content.}
\end{table*}

\begin{table}[ht!]  
\tablenum{5}
\begin{center}
\center{\caption{Unblended photospheric lines in O9.5--O4 spectra (non-Fe\,V)}}
\begin{tabular}{rllllll}
\tableline\tableline
Ion & Exc. &  &  &  &  &  \\
Wavel & O9.5~~~~ & O8 & O7 & O6 & O5 &  O4 \\
\tableline
He\,II~~~~~ & [40.8] & & & & & \\
 958.698 &  1  &  2  &  3  & 4  &  5  &  6 \\
 972.111 &  1  &  2  &  3  & 4  &  5  &  6 \\
 992.363 &  1  &  2  &  3  & 4  &  5  &  6 \\
1025.272 &  1  &  2  &  3  & 4  &  5  &  6 \\
1084.942 &  1  &  2  &  3  & 4  &  5  &  6 \\
&  &  &  & &  &  \\
C\,II~~~~~~ & [5.3 eV] & & & & & \\
1010.371 &  1  &  2  &  3  & 4  &     &    \\
&  &  &  & &  &  \\
C\,III~~~~~ & [6.5--33.5 eV] & & & & & \\
977.020  &  1  &  2  &  3  &  4 & 5   & 6 \\
1070.331 &  1  &  2  &  3  & 4  &  5  &    \\
1125.669 &  1  &  2  &  3  & 4  &  5  &  6 \\
1139.899 &  1  &  2  &  3  & 4  &  5  &    \\
1165.623 &  1  &  2  &  3  & 4  &  5  &  6 \\
1174.933 &  1  &  2  &  3  & 4  &  5  &    \\
1175.263 &  1  &  2  &  3  & 4  &  5  &    \\
1175.590 &  1  &  2  &  3  & 4  &  5  &  6 \\
1175.711 &  1  &  2  &  3  & 4  &  5  &    \\
1175.987 &  1  &  2  &  3  & 4  &  5  &  6 \\
1176.369 &  1  &  2  &  3  & 4  &  5  &  6 \\
&  &  &  & &  &  \\
C\,IV~~~~~ & [39.7--49.7] & & & & & \\
1097.319 &  1  &   2  &  3 & 4  &  5  &  6 \\
1107.591 &  1  &   2  &  3  &  4  &  5  &    \\
1107.930 &  1  &   2  &  3  &  4  &  5  &    \\
1168.847 &  1  &   2  &  3  &  4  &  5  &  6 \\
1168.990 &  1  &   2  &  3  &  4  &  5  &  6 \\
&  &  &  & &  &  \\
N\,III~~~~~ & [17.2--33.1] & & & & & \\
1103.044 &  1  &   2  &   3 &  4  &     &    \\
1106.036 &  1  &   2  &   3 &  4  &  5  &    \\
1106.340 &  1  &   2  &   3 &  4  &     &    \\
1112.648 &  1  &   2  &   3 &  4  &  5  &  6 \\
1132.883 &  1  &   2  &   3 &  4  &  5  &    \\
1135.762 &  1  &   2  &   3 &  4  &     &    \\
1140.104 &  1.2  &   2.2  &  3.2 &  4.2  &     &    \\
1152.406 &  1  &   2  &   3 &  4  &     &    \\
1152.627 &  1  &   2  &   3 &  4  &     &    \\
1183.032 &  1.1  &   2.1  &   3.1 &  4.1  &     &    \\
1184.514 &  1.1  &   2.1 &   3.1 &  4.1  &     &    \\
&  &  &  & &  &  \\
\tableline
\end{tabular}
\end{center}
\end{table}

\begin{table}[ht!]  
\tablenum{5}
\begin{center}
\center{\caption{Unblended photospheric lines in O9.5--O4 spectra (non-Fe\,V)}}
\begin{tabular}{rllllll}
\tableline\tableline
Ion & Exc. &  &  &  &  &  \\
Wavel & O9.5~~~~ & O8 & O7 & O6 & O5 &  O4 \\
\tableline
N\,IV~~~~~ & [16.9--53.2] & & & & & \\
 955.334 &  1  &   2  &   3 &  4  &   5 &  6 \\
1004.222 &  1  &   2  &   3 &  4  &   5 &    \\
1078.711 &  1  &   2  &   3 &  4  &   5 &    \\
1131.488 &  1  &   2  &   3 &  4  &   5 &  6 \\
1133.121 &  1  &   2  &   3 &  4  &   5 &    \\
1135.252 &  1  &   2  &   3 &  4  &   5 &    \\
1135.485 &  1  &   2  &   3 &  4  &   5 &    \\
1142.151 &  1  &   2  &   3 &  4  &   5 &    \\
1163.264 &  1  &   2  &   3 &  4  &   5 &    \\
1173.018 &  1  &   2  &   3 &  4  &   5 &    \\
1188.005 &  1  &   2.1 & 3 &  4  &   5 &  6 \\
&  &  &  & &  &  \\
O\,III~~~~~ & [5.4--38.0]  & & & & & \\
 962.425 &  1  &   2  &   3 &  4  &   5 &  6 \\
1007.875 &  1  &   2  &   3 &  4  &   5 &  6 \\
1008.097 &  1  &   2  &   3 &  4  &   5 &  6 \\
1016.717 &  1  &   2  &   3 &  4  &   5 &  6 \\
1020.415 &  1  &   2  &   3 &  4  &   5 &    \\
1020.549 &  1  &   2  &   3 &  4  &   5 &    \\
1020.780 &  1  &   2  &   3 &  4  &   5 &    \\
1022.280 &  1  &   2  &   3 &  4  &   5 &    \\
1031.493 &  1  &   2  &   3 &  4  &   5 &  6 \\
1040.320 &  1  &   2  &   3 &  4  &   5 &  6 \\
1050.358 &  1  &   2  &   3 &  4  &   5 &    \\
1056.373 &  1  &   2  &   3 &  4  &   5 &    \\
1056.433 &  1  &   2  &   3 &  4  &   5 &    \\
1058.674 &  1  &   2  &   3 &  4  &   5 &    \\
1059.005 &  1  &   2  &   3 &  4  &   5 &    \\
1082.020 &  1  &   2  &   3 &  4  &   5 &    \\
1098.473 &  1  &   2  &   3 &  4  &   5 &    \\
1138.551 &  1  &   2  &   3 &  4  &   5 &  6 \\
1149.602 &  1  &   2  &   3 &  4  &   5 &  6 \\
1150.882 &  1  &   2  &   3 &  4  &   5 &  6 \\
1153.022 &  1  &   2  &   3 &  4  &   5 &    \\
1153.775 &  1  &   2  &   3 &  4  &   5 &    \\
1157.041 &  1  &   2  &   3 &  4  &   5 &  6 \\
1157.161 &  1  &   2  &   3 &  4  &   5 &  6 \\
1160.154 &  1  &   2  &   3 &  4  &   5 &  6 \\
1172.451 &  1  &   2  &   3 &  4  &   5 &  6 \\
&  &  &  & &  &  \\
\tableline
\end{tabular}
\end{center}
\end{table}

\begin{table}[ht!]  
\tablenum{5}
\begin{center}
\center{\caption{Unblended photospheric lines in O9.5--O4 spectra (non-Fe\,V)}}
\begin{tabular}{rllllll}
\tableline\tableline
Ion & Exc. &  &  &  &  &  \\
Wavel & O9.5~~~~ & O8 & O7 & O6 & O5 &  O4 \\
\tableline
O\,IV~~~~~ & [48.4--63.4] & & & & & \\
 988.708 &     &   2  &   3 &  4  &   5 &  6 \\
1045.364 &  1  &   2  &   3 &  4  &   5 &    \\
1046.313 &  1  &   2  &   3 &  4  &   5 &    \\
1067.768 &  1  &   2  &   3 &  4  &   5 &  6 \\
1067.958 &  1  &   2  &   3 &  4  &   5 &    \\
1080.969 &  1  &   2  &   3 &  4  &   5 &  6 \\
1164.546 &  1  &   2  &   3 &  4  &   5 &    \\
1167.478 &  1  &   2  &   3 &  4  &   5 &    \\
&  &  &  & &  &  \\
O\,V~~~~~~~ & [92.0--92.5] &  & & & & \\
 995.087 &     &   2  &   3 &  4  &   5 &  6 \\
1010.602 &  1  &   2  &   3 &  4  &   5 &  6 \\
1090.320 &  1  &   2  &   3 &  4  &   5 &  6 \\
&  &  &  & &  &  \\
O\,V~~~~~~~ & [.0.0] &  & & & & \\
1031.926 &  1.1 &  2.1 &  3  &  4  &  5.1 &  6.1 \\
1037.617 &  1  &  2  &  3  &  4  &  5  &  6 \\
&  &  &  & &  &  \\
Si\,III~~~~~ & [6.5--6.6]  &     &     &     &    \\
1108.358 &  1  &   2  &   3 &  4  &   5 &    \\
1109.970 &  1  &   2  &   3 &  4  &   5 &    \\
1113.204 &  1  &   2  &   3 &  4  &   5 &  6 \\
&  &  &  & &  &  \\
Si\,IV~~~~~ & [8.8--31.0]  &     &     &     &    \\
1066.614 &  1  &   2  &   3 &  4  &   5 &  6 \\
1122.485 &  1  &   2  &   3 &  4  &   5 &  6 \\
1154.621 &  1  &   2  &   3 &  4  &   5 &  6 \\
&  &  &  & &  &  \\
P\,IV~~~~~ & [0.0--27.2]  &     &     &     &    \\
 950.657 &  1  &   2  &   3 &  4  &   5 &  6 \\
1030.515 &  1.1  &   2.1  &  3.1 & 4.1  &  5.1 & 6.1 \\
1088.616 &  1  &   2  &   3 &  4  &   5 &  6 \\
1091.442 &  1  &   2  &   3 &  4  &   5 &    \\
1118.552 &  1  &   2  &   3 &  4  &   5 &  6 \\
1187.540 &  1  &   2  &   3 &  4  &   5 &    \\
&  &  &  & &  &  \\

P\,V~~~~~ & [0.0]  &     &     &     &    \\
1000.358  &  1  &   2 & 3  & 4  & 5  &  6 \\
1118.043 &  1  &   2  &   3 &  4  &   5 &  6 \\
1128.010 &  1  &   2  &   3 &  4  &   5 &  6 \\
&  &  &  & &  &  \\
\tableline
\end{tabular}
\end{center}
\end{table}

\begin{table}[ht!]  
\tablenum{5}
\begin{center}
\center{\caption{Unblended photospheric lines in O9.5--O4 spectra (non-Fe\,V)}}
\begin{tabular}{rllllll}
\tableline\tableline
Ion & Exc. &  &  &  &  &  \\
Wavel & O9.5~~~~ & O8 & O7 & O6 & O5 &  O4 \\
\tableline
S\,IV~~~~~~ & [0.0--22.5] & & & & & \\
1062.662 &  1  &   2  &   3 &  4  &   5 &  6 \\
1072.974 &  1  &   2  &   3 &  4  &   5 &  6 \\
1073.516 &  1  &   2  &   3 &  4  &   5 &  6 \\
1098.362 &     &   2  &   3 &  4  &   5 &  6 \\
1098.929 &  1  &   2  &   3 &  4  &   5 &  6 \\
1099.482 &  1  &   2  &   3 &  4  &   5 &  6 \\
1100.051 &  1  &   2  &   3 &  4  &   5 &  6 \\
1106.487 &  1  &   2  &   3 &  4  &   5 &  6 \\
1107.733 &  1  &   2  &   3 &  4  &   5 &  6 \\
1108.449 &  1  &   2  &   3 &  4  &   5 &  6 \\
1110.905 &  1  &   2  &   3 &  4  &   5 &  6 \\
1111.044 &  1  &   2  &   3 &  4  &   5 &  6 \\
1111.253 &  1  &   2  &   3 &  4  &   5 &  6 \\
1117.161 &  1  &   2  &   3 &  4  &   5 &  6 \\
1138.076 &  1  &   2  &   3 &  4  &     &     \\
1138.210 &  1  &   2  &   3 &  4  &  5  &   6 \\
&  &  &  & &  &  \\
Fe\,III~~~~~& [16.7] &  &  & &  &  \\
1091.082   & 1  & 2 & 3 & 4 & &    \\
&  &  &  & &  &  \\
Fe\,IV~~~~~& [17.2-20.5] &  &  & &  &  \\
1005.697 &  1   &   2  &   3   &  4  &   5  &   \\
1126.625 &  1   &   2  &   3   &  4  &   5  &   \\
1131.033 &  1   &   2  &   3   &  4  &   5  &   \\
1156.526 &  1   &   2  &   3   &  4  &   5  &   \\
1160.561 &  1   &   2  &   3   &  4  &      &   \\
1170.781 &  1   &   2  &   3   &  4  &      &   \\
1174.121 &  1   &   2  &   3   &  4  &      &   \\
1176.973 &  1   &   2  &   3   &  4  &      &   \\
&  &  &  & &  &  \\
\tableline

Ion & Exc. &  &  &  &  &  \\
Wavel & O9.5~~~~ & O8 & O7 & O6 & O5 &  O4 \\
\tableline
Fe\,VI~~~~~& [34.9-36.2] &  &  & &  &  \\
1160.509   &    &   &    &   4 &    5  &   6 \\
1170.275   &    &   &    &   4 &    5  &   6 \\
1186.575   &    &   &    &   4 &    5  &   6 \\
\tableline
\end{tabular}
\end{center}
\end{table}

\begin{table}[ht!]
\tablenum{6}
\begin{center}
\center{\caption{Unblended photospheric lines in O9.5-O4 spectra (Fe\,V)}}
\begin{tabular}{rllllll}
\tableline\tableline
Ion & Exc.  &  &  &  &  &   \\
Wavel & O9.5~~~~ & O8 & O7 & O6 & O5 &  O4 \\
\tableline
Fe\,V~~~~~& [23.1-41.7]  &  &  &  &  & \\
 951.073  & -   &   2   &  3   &  4   &  5   & - \\
 951.294  & -   &   2   &  3   &  4   &  5   & - \\
 951.449  & -   &   -    &  3   &  4   &  5   & -  \\
 953.847  & 1   &   2   &  3   &  4   &  5   &  6 \\
 954.534  & 1   &   2   &  3   &  4   &  5   &  6 \\
 956.879  & -   &   2   &  3.1 &  4.1 &  5   &  6 \\
 957.225  & 1.1 &   2.1 &  3.1 &  4.1 &  5.1 &  6.1 \\
 957.909  & 1   &   2   &  3   &  4   &  5   &  6 \\
 958.097  & 1.1 &   2.1 &  3.1 &  4.1 &  5.1 &  6.1 \\
 958.288  & -   &   2   &  3   &  4   &  5   &  6 \\
 958.379  & -   &   -   &  3   &  4   &  5   &  6 \\
\tableline
\end{tabular}
\end{center}
\tablecomments{Table\,6 is presented in its entirety
in the MAST archives at http://archive.stsci.edu/prepds/fuvostars/.
A portion is shown here for guidance in data format and content.}
\end{table}

\end{document}